# Scintillation response of $Ga_2O_3$ excited by laser accelerated ultra-high dose rate proton beam

Yulan Liang[1], Tianqi Xu[1], Shirui Xu[1], Qingfan Wu[1], Chaoyi Zhang[2], Haoran Chen[1], Qihang Han[1], Chenhao Hua[1], Jianming Xue[1], Huili Tang[2, *], Bo Liu[2, *] and Wenjun Ma[1, *]

[1] State Key Laboratory of Nuclear Physics and Technology, School of Physics, Peking University, Beijing 100871, P. R. China;

[2] MOE Key Laboratory of Advanced Micro-Structure Materials, School of Physics Science and Engineering, Tongji University, Shanghai 200092, China;

*Corresponding author. Email: wenjun.ma@pku.edu.cn; lbo@tongji.edu.cn

**Abstract**:

The temporal and spectral profile of $β-Ga_2O_3$ excited by ultra-high dose rate proton beam has been investigated. The unique short bright and broad spectra characteristics of laser-accelerated protons were utilized to investigate the scintillation response difference under different dose rate. Our results indicate that for sufficiently high dose rate delivered, the average decay time of $β-Ga_2O_3$ decreases by a factor of two. The overlap of carriers generated by high dose rate protons enhances the nonradiative recombination like Auger recombination and exciton-exciton annihilation which shortens the decay time significantly. The study opens up new avenues for investigating the luminescent properties of other scintillator materials using laser-accelerated high dose rate proton beams.

**(Introduction)**

The development of high dose rate radiation not only promote scientific progress but also plays a crucial role in energy development, human health and environmental protection, encompassing a wide range of applications, including advanced FLASH radiotherapy, inertial confinement fusion, nuclear energy, and high energy physics.[1, 2] Radiation diagnosis and evaluation of these fields require high-performance scintillation detectors that exhibit strong radiation resistance, fast response time, and relatively high light yield.[3] The response of many commercial detectors exhibits nonlinear behavior with respect to the energy loss per unit distance, necessitating consideration of carrier quenching effects when utilizing these detectors in high dose rate irradiation.[4, 5] Mechanism studies indicate that the increase in excitation density can give rise to processes such as exciton-exciton annihilation, defect capture, and Auger recombination, which in turn lead to nonlinear changes in the luminescent properties of the materials.[6] Investigations have demonstrated nonlinear variations in luminescent light yield and lifetime in materials such as BGO, $CdWO_4$, $WSe_2$, and novel perovskite materials, with changes in laser power density.[6-10] Therefore, it is of paramount importance to characterize and investigate materials under high dose rate and high fluence rate application scenarios before practical applications.

However, the current characterization of scintillation detectors uses techniques such as laser, electron gun, X-ray tubes, various radioactive isotopes, and some linear accelerators.[3, 11, 12] They can merely focuses on the parameter range associated with low fluence rate and low dose rate. Moreover, present ongoing research on nonlinear effects primarily relies on high energy femtosecond (fs) pulse lasers.[13] The fundamental processes of high energy particle excitation and photoluminescence excitation differ significantly.[14, 15] Optical absorption leads to gradual proliferation of electron-hole pairs. Unlike through band excitation, high energy particles interact mainly with the electrons of the scintillators, resulting in ionized secondary electrons along the particle tracks with energy up to keV. These ionized secondary electrons form localized spur with electron-hole pair more densely populated, which in turn complicates the physical process.[16, 17] Due to limitations associated with ion beam accelerators, there is little reports on related nonlinear studies. The majority of research in this domain has concentrated on the low-fluence-rate range and been confined to scrutinizing the discrepancies in luminescence induced by different particle types.[18, 19]

Rapid development of laser plasma accelerators attracts attention as a compact alternative to conventional accelerators with high-energy, high-peak proton beams.[20, 21] The short, bright unique characteristics serves as a useful potential tool to extend our ability to study scintillation properties under high dose rate beam irradiation. In this study, we made use of the broad spectrum and short pulse width features of laser-accelerated proton beams. As a novel approach, we employed ultra-high dose rate protons up to $8.73 \times 10^{11}$ Gy/s. Subsequently, we conducted a systematic investigation into the luminescence lifetime and spectral characteristics of β-Ga$_2$O$_3$ under various dose rate proton excitations.

**(Results and Discussion)**

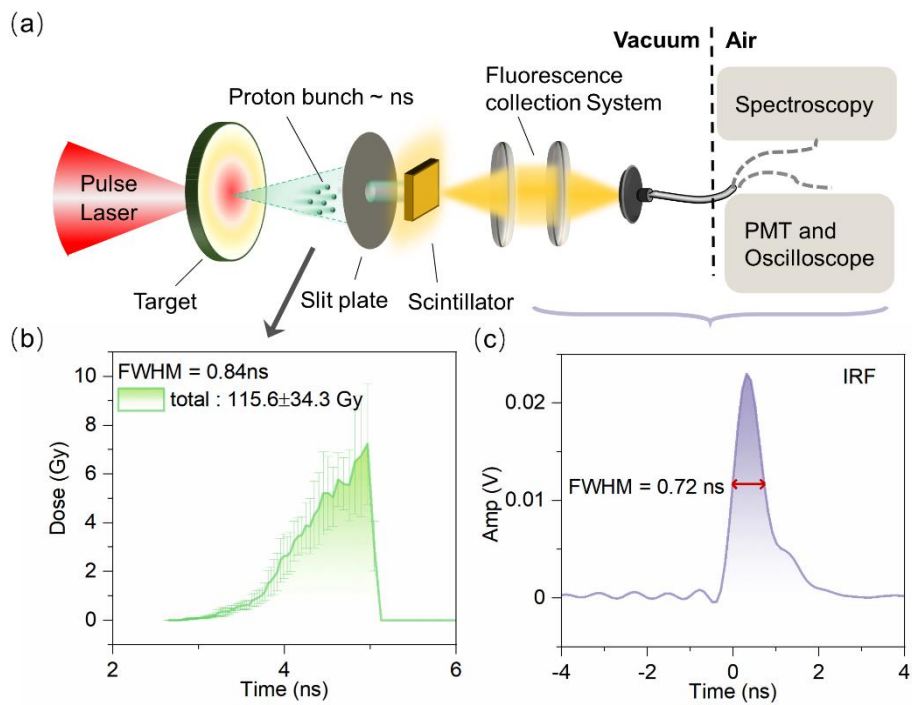

**FIG. 1**. System information for UHDR excitation. (a) Schematic diagram of scintillation detection excited by proton beam at CLAPA system. (b) Dose deposited in β-Ga$_2$O$_3$ when placed 0.085m from the proton pulse. (c) IRF of the whole detecting system, including the Cage System and the corresponding Electronic component system.

A complex platform for ultra-high dose rate (UHDR) proton excited luminescence investigation, both temporal and spectral, were built and added to the Compact Laser Plasma Accelerator (CLAPA) in Peking University (**Fig. 1a**). Laser accelerated protons are generated in a picosecond, with an exponentially decaying proton energy spectrum from ~10$^{11}$ protons/MeV/sr at 1 MeV to ~10$^{8}$ protons/MeV/sr at 4 MeV (**Fig. S1**). Protons with energies below 1 MeV are unable to penetrate the aluminum foil that serves as a laser blocking barrier before reaching the scintillator. Moreover, in comparison to 1 MeV protons, the population of 2 MeV protons has already decreased by an order of magnitude. Therefore, excitation process mainly concentrates on ions within the 1–2 MeV range with relatively minimal spatial and temporal dispersion. By incorporating the measurements of actual dose by radiochromic film (RCF) and Thomson Parabola spectrometer (TPS) measured spectrum into FLUKA (FLUktuierende KAskade) simulations, we can derive that at a distance of 4.6 cm from the target point, the scintillator receives a total of $6.5\times10^9$ protons, with a maximum energy deposition of $6.9\times10^{10}$ MeV in nanosecond (ns) scale.[22] The precise adjustment of the irradiation dose rate can be achieved by manipulating the position of the scintillator relative to the proton beam outlet, taking into account the divergence of the beam. Even when considering the flight time of the overall energy spectrum, the full width at half maximum (FWHM) of the beam remains within 1 ns for all experiments (**Fig S2**), predominantly influenced by the rising edge. This change exhibits a steep falling edge compared to the lifetimes of β-$Ga_2O_3$, which range from tens to hundreds of ns, rendering it negligible for practical considerations. The energy spectrum is repeatedly tested before and after scintillation measurements, with the averaged dose rate demonstrating stability of the proton beam (**Fig 1b**). The complete setup encompasses fluorescence collection, fiber optic transmission, photomultiplier tubes, and an oscilloscope. The impulse response function (IRF) information of the entire system is maintained within 2 ns (**Fig. 1c**), obtained via a 400nm femtosecond pulse laser.

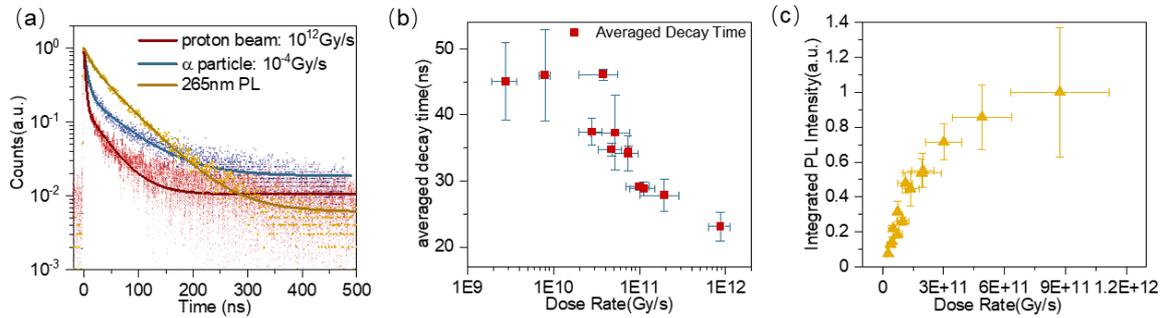

**FIG. 2**. Comparison of fluorescence temporal profile of β-$Ga_2O_3$ under different radiation sources. (a) The fluorescence profile under Photoluminescence, and proton beam of LDR and UHDR. (b) (c) The averaged decay time and normalized light yield under proton excitation of different dose rate. Each point has been repeated at least three times with the y-axis error-bar being the standard deviation among repeated shots. The x-axis error-bar were reported from the fluctuation of the proton beam source.

TRPL (Time-Resolved Photoluminescence) measurement was carried out using 265nm pulsed laser. By fitting the decay curve, fast/slow components of 12.7/57.2 ns were obtained, consistent with previous studies (**Fig. 2a**).[23-25] The average lifetime ($\tau_{ave}$) is 52.9 ns. Following this, the temporal response of β-$Ga_2O_3$ to proton excitation at varying dose rates was meticulously examined. In order to more effectively illustrate the impact of dose rate on lifetime, both UHDR proton and low-dose-rate (LDR) $\alpha$-particle source excitations were conducted with a significant disparity in fluence rate values. The UHDR experiments were performed in CLAPA, with maximum dose rate up to $10^{12}$ Gy/s. On the other hand, the LDR experiments were conducted using radioactive isotopes $^{241}$Am (5.52 MeV) with dose rate merely $10^{-4}$Gy/s. For comparison convenience, both data are fitted with bi-exponential function for time range 0-500ns (**Fig. 2a**). The fast/slow component is 2.9/36.5 ns for UHDR, and 6.4/64.9 ns for LDR. The $\tau_{ave}$ of β-Ga$_2$O$_3$ are derived to be 26.6/51.6 ns for UHDR/LDR excitations, which shows a twofold decrease.

The average lifetime does not show significant difference for TRPL and LDR. However, the LDR curve displayed an obvious long tail and initial faster decay compared to PL. This could be resulted from the difference of excitation nature. In contrast to UV light excitation of band-edge electrons, ionoluminescence excited electrons to much higher energy levels, which undergo more complex procedures of thermalizations and transportations. A densely localized electron-hole pair can result in Auger recombination, exciton-exciton annihilation and more defect capture procedure which change the temporal shape.[16, 26] As UHDR exceeds LDR by as high as sixteen magnitudes, $\tau_{ave}$ shortened approximately by a factor of two. For UHDR irradiation, the carrier concentration within the scintillator will undergo even more rapid and significant increase, which can result in more carrier quenching and carrier scatterings. Thereby, shortening its temporal properties.

To better elucidate the impact of proton dose rate on the lifetime, a rigorous investigation was conducted in CLAPA. β-Ga$_2$O$_3$ was exposed to pulsed proton beam with dose rate ranging from $10^{10}$Gy/s to $10^{12}$Gy/s. **Fig. 2b** summarizes average lifetimes for each dose rate and each point has been repeated at least 3 times to obtain standard deviation. When the dose rate decreases to approximately 2.8×$10^{10}$Gy/s, the average lifetime of the scintillator aligns with that of LDR and TRPL, differing by no more than 10%. With increasing dose rate, the overall lifetime experienced a precipitous decrease. Upon reaching a dose rate of 8.73×$10^{11}$Gy/s, the lifetime reaches 20 ns, which is 2–3 times smaller compared to LDR and TRPL. This serves to underscore the non-linear correlation between the scintillation lifetime of β-Ga$_2$O$_3$ and the dose rate of proton excitation. The experimental data exhibits a turning point somewhere around $10^{10}$Gy/s, where the lifetime of the proton-irradiated material shows no significant change before, but experiences a notable decrease thereafter. As the carrier concentration escalates to a certain region, more intricate physical processes necessitate contemplation. X-ray diffraction (XRD) and PL were taken before and after proton irradiation, suggesting crystal structure stability and no obvious scintillation response changes for the material(**Fig. S3/S4**). Therefore, the lifetime difference observed during experiment is likely the cause of changes in dose rate, possibly the result of non-radiative quenching. The light yield exhibits a nonlinear response with increasing dose, reaching a platform after an initial linear enhancement (**Figure 2d**). Such a nonlinear dependence reveals that luminescence recombination channels is close to a saturation with increasing dose rate. Higher dose rate generates larger carrier concentration. We deduce that with a dense carrier concentration, some higher-order processes like exciton-exciton annihilation and Auger recombination can enlarge the nonradiative recombination rates and become more effective.[10, 27-29]

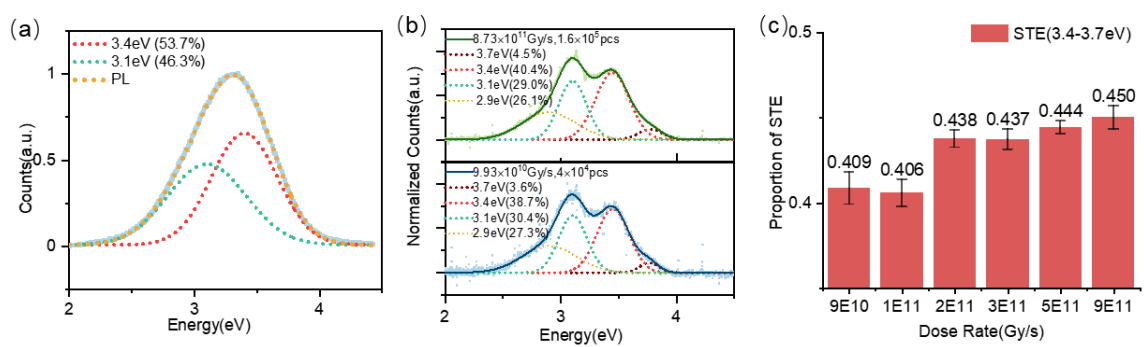

**Figure 3**. Comparison of luminescence spectra of β-$Ga_2O_3$ under different excitation. (a) 265nm excitation photoluminescence. (b) Proton excitation luminescence under different dose rate with the proportion of self-trapped-exciton luminescence slightly changes with dose rate shown in (c).

In order to further understand the lifetime changes, the luminescence spectra under different radiation are then carefully investigated. The Photoluminescence (PL) spectra with 265nm excitation are shown in **Fig. 3a**, fitted by two Gaussians in UV/ UV' band.[30-32] Both bands contributed roughly equally. The UV (3.2-3.7 eV) emission is attributed to the recombination of self-trapped hole (STH) and free electrons, whereas the UV' (3.1 eV) luminescence arise from donor-acceptor pair recombination.[30, 33]

To shed light on the excitation mechanisms behind the temporal difference excited by pulsed proton beam under different dose rate, in-situ luminescence spectra measurements were carried out (**Fig. 3b**). Compared with PL, UHDR spectra are composed with four peaks, 3.7eV $UV_1$, 3.4eV $UV_2$(~40%), 3.1eV UV' (~30%), 2.9eV BL (~25%). Large FWHM leaded to serious overlaps, indicating a strong phonon-electron coupling. Both $UV_1$ and $UV_2$ emission are the recombination of STH and free electrons. Both theoretically and experimentally reported, holes has the tendency to form polarons with lattice distortion in two locations : O(I) 3.7eV and dual O(II) 3.2-3.6eV.[34] The appearance of the BL suggests that deeper defects are involved in the radiative recombination luminescence process of proton-excited fluorescence. For proton irradiation, electrons are excited to much higher energy levels which undergo inelastic scatterings and thermalization before recombination. Therefore, the lattice distortion is relatively weak compared to 265nm laser excitation. Electrons can transfer directly to defects levels and contribute to lower energy radiative recombination. Similar difference between radioluminescence and PL for $Cs_3Cu_2I_5$ has been observed.[26]

The ratio of UV ($UV_1$ and $UV_2$) shows an initial increase trend with higher dose-rate and reaches a platform after $2\times10^{11}$Gy/s (**Fig. 3c**). UV luminescence usually leads to a faster decay time compared with donor-acceptor-pair recombination.[35] Therefore, the slight increase ratio of UV luminescence can result in a faster averaged decay time. However, the temporal profile continues to be shortened while the proportion of UV luminescence reached a platform. Moreover, the overall luminescence spectra do not show any significant changes (**Fig. 3c**) with the averaged lifetime undergone a 50% reduction (**Fig. 2b**) during the same dose rate difference. This indicates that the ratio between two luminescence mechanisms stays relatively stable. A higher dose-rate would result in higher carrier density and localized temperature. Consequently more non-radiative quenching including trap capture, Auger recombination and exciton-exciton annihilation can take place which shortens the averaged decay time significantly.

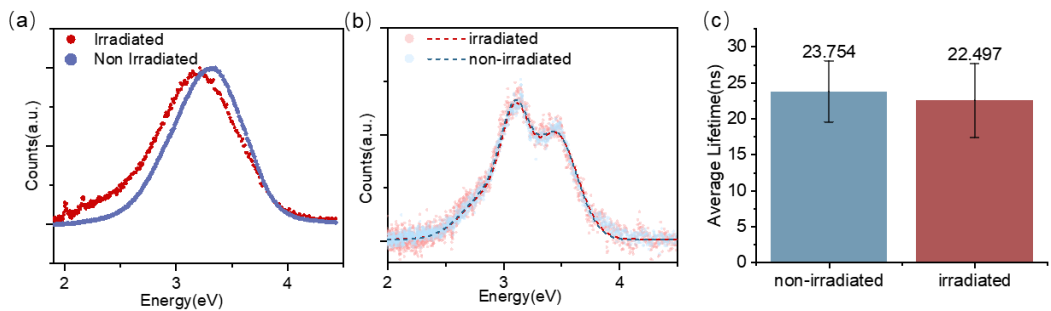

**Figure 4**. Comparison of scintillation properties of $Ga_2O_3$ before and after $10^{15}$ cm$^{-2}$ proton irradiation. (a) 265nm excitation PL. (b) UHDR proton excited luminescence spectra. (c) UHDR proton excited averaged lifetime.

To further investigate the non-radiative quenching during UHDR excitation, the same β-$Ga_2O_3$ sample has been irradiated by 2MeV proton beam with $10^{15}cm^{-3}$ fluence on purpose. The PL spectra shows an increase in low energy luminescence (**Fig. 4a**), indicating more defects have been generated. However, the in-situ temporal and spectra profile does not show a significant difference under UHDR irradiation before and after the irradiation (**Fig. 4b/4c**). The differences in defects concentration do not influence scintillation response in UHDR excitation. This strongly support that, during UHDR irradiation, higher-order processes like exciton-exciton annihilation and Auger recombination can enlarge the nonradiative recombination rates and the defects concentration difference is negligible.

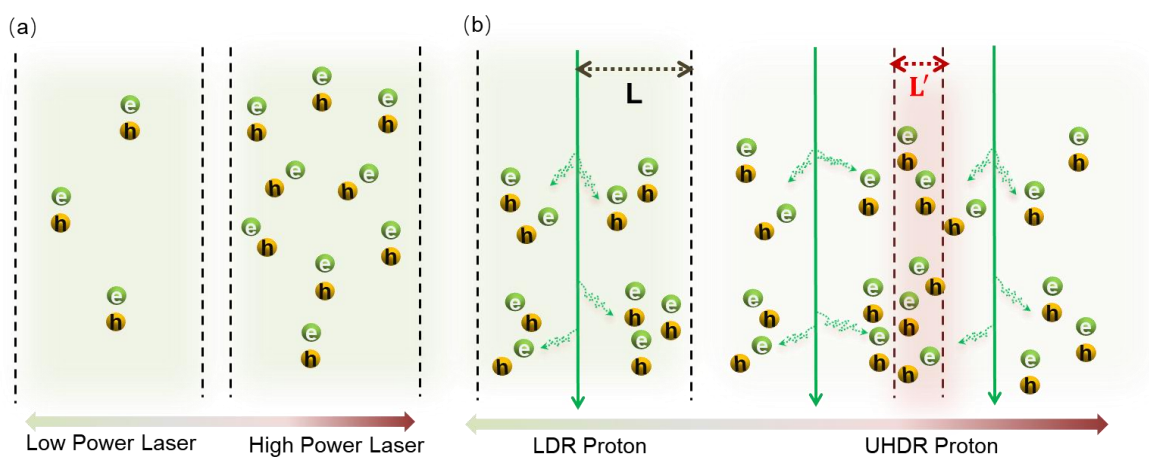

**Figure 5**. Photon, LDR proton and UHDR proton excitation Difference.

Lasers are typically used to uniformly excite sample luminescence. By increasing the optical power density, the overall carrier concentration in the irradiated region of the material can be enhanced (**Fig. 5a**). To achieve carrier concentrations in the range of nonlinear effects, high fluence lasers with mJ/cm² are required.[7] Moreover, for PL, carriers are generated either by through band or below band excitation. Electrons are mainly distributed near the valence band. However, the situation is different for proton excitation. Even a single proton can generate high-energy electrons through ionization energy loss, resulting in a high concentration of carriers locally. The energy loss of 1.8 MeV protons along the path is about 100 eV/nm, mainly distributed within a radius of 100 nm (**Fig. S5**). The carrier concentration formed in this localized space can reach as high as $10^{19}cm^{-3}$. Therefore, even for a single proton, the localized high carrier concentration can lead to more carrier dynamics in the luminescence process. Donor-acceptor luminescence typically involves a tunnel capture process. High-energy electrons and the locally high carrier concentration can induce localized heating, thereby enhancing the probability of tunnel capture. As a result, the defect luminescence peak under proton excitation is significantly enhanced compared to photoluminescence.

As the dose rate of incident protons increases, carriers generated by individual protons during the diffusion process may overlap, further enhancing the local carrier concentration (**Fig. 5b**). The diffusion length of β-$Ga_2O_3$ is approximately 400 nm.[36] When the dose rate reaches $8.73 \times 10^{11}$ Gy/s, the corresponding proton concentration is ~20 protons/(400nm)$^2$. Therefore, the electrons and holes generated by protons will highly overlap. For the material, the limited defect concentration makes the intensity of defect luminescence more prone to saturation compared to self-trapped exciton (STE) luminescence. Therefore, with increasing dose rate, the contribution of STE shows a slight increase. The lack of significant changes in the overall spectral shape with dose rate indicates that the additional excess carriers at this point participate in non-radiative recombination processes, like Auger recombination and exciton-exciton annihilation accelerating the carrier lifetime.[16] The saturation of luminescence at high dose rates also reflects the presence of a significant amount of non-radiative recombination.

**(Conclusion)**

In summary, this study presents a systematic investigation of the luminescence lifetime and spectral characteristics of β-$Ga_2O_3$ under various dose rate of proton excitations. For the first time, we examine the effects of proton beam dose rate on scintillation response of β-$Ga_2O_3$. It is observed that when the dose rate exceeds $10^{10}$ Gy/s, the averaged lifetime decreases significantly by a factor of two. However, no significant changes are observed in the emission spectrum. This decrease in lifetime can be attributed to the enhanced nonradiative processes, which becomes more dominant with increasing dose rate. Beyond the threshold of $10^{10}$ Gy/s, the effects on β-$Ga_2O_3$ are no longer solely dependent on individual proton interactions. The overlap of carriers generated between protons enhances the nonradiative recombination, resulting in a reduction in the luminescence lifetime. Furthermore, the study reveals that defect concentration has minimal impact on high dose rates proton excitation response for the material, indicating a higher tolerance for material quality under such conditions. These findings establish the potential application of β-$Ga_2O_3$ in the field of high energy physics. Moreover, the study opens up new avenues for investigating the luminescent properties of other scintillator materials using laser-accelerated high dose rate proton beams.